\documentclass[runningheads]{llncs}

\usepackage{graphicx}
\usepackage{microtype}
\usepackage[caption=false]{subfig} 

\usepackage{makecell}  
\usepackage{xcolor,colortbl}

\definecolor{Gray}{gray}{0.85}
\definecolor{LightCyan}{rgb}{0.88,1,1}
\newcolumntype{a}{>{\columncolor{Gray}}l}
\newcolumntype{b}{>{\columncolor{LightCyan}}l}

\newcolumntype{T}{>{\tiny}c} 

\usepackage{booktabs} 

\usepackage{amsmath}
\usepackage{microtype}
\usepackage{multirow}

\usepackage[acronym]{glossaries}
\newacronym{TOPIC}{TOPIC}{Challenges and complexities in Machine Learning based Credit Card Fraud Detection}

\begin{document}

\title{Challenges and Complexities in Machine Learning based Credit Card Fraud Detection}
\titlerunning{Challenges and Complexities in ML based CC Fraud Detection}

\author{Gayan K. Kulatilleke}
\authorrunning{G. K. Kulatilleke}
%
\institute{Work outlined in this paper is part of the author’s MSc dissertation at Queen Mary University of London, 2017.\\\email{tidalbobo@gmail.com}}
%
\maketitle  
\begin{abstract}
Credit cards play an exploding role in modern economies. Its popularity and ubiquity have created a fertile ground for fraud, assisted by the cross boarder reach and instantaneous confirmation. While transactions are growing, the fraud percentages are also on the rise as well as the true cost of a dollar fraud. Volume of transactions, uniqueness of frauds and ingenuity of the fraudster are main challenges in detecting frauds. The advent of machine learning, artificial intelligence and big data has opened up new tools in the fight against frauds. Given past transactions, a machine learning algorithm has the ability to 'learn' infinitely complex characteristics in order to identify frauds in real-time, surpassing the best human investigators. 
However, the developments in fraud detection algorithms has been challenging and slow due the massively unbalanced nature of fraud data, absence of benchmarks and standard evaluation metrics to identify better performing classifiers, lack of sharing and disclosure of research findings and the difficulties in getting access to confidential transaction data for research. 
This work investigates the properties of typical massively imbalanced fraud data sets, their availability, suitability for research use while exploring the widely varying nature of fraud distributions. 
Furthermore, we show how human annotation errors compound with machine classification errors.
We also carry out experiments to determine the effect of PCA obfuscation (as a means of disseminating sensitive transaction data for research and machine learning)  on algorithmic performance of classifiers and show that while PCA does not significantly degrade performance, care should be taken to use the appropriate principle component size (dimensions) to avoid overfitting.

\end{abstract}

\section{Introduction} \label{Introduction}
Credit cards play a major role in today's economy \cite{zojaji2016survey}. Card fraud includes any theft or fraud involving a credit or debit payment card. The perpetrator could be the authorized user attempting an illicit activity, or an unauthorized user committing a fraud by means of a stolen card or compromised card credentials. Financial Fraud Action UK (FFA UK) reports 19.1 billion transactions during 2016 in UK, a 10\% increase from previous year and “fraud losses totalling £618.0 million, a 9\% increase from the previous year”, its fifth consecutive year of increases. Current prevention rates are only 0.6/£.

\begin{figure}
    \centering \includegraphics[width=1.00\columnwidth]{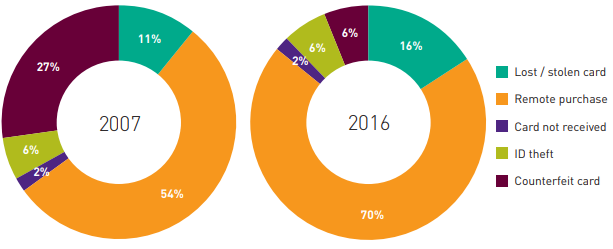}
    \caption{Losses by type of Fraud, as a \% of total loss. 
    \\Source: www.financialfraudaction.org.uk/fraudfacts17/assets/fraud\_the\_facts.pdf}
    \label{fig_CC_losses}
\end{figure}

Figure~\ref{fig_CC_losses} shows that majority of the frauds are remote purchases, carried out online via the Internet, where the card is not present. 

\begin{figure}
    \centering \includegraphics[width=1.00\columnwidth]{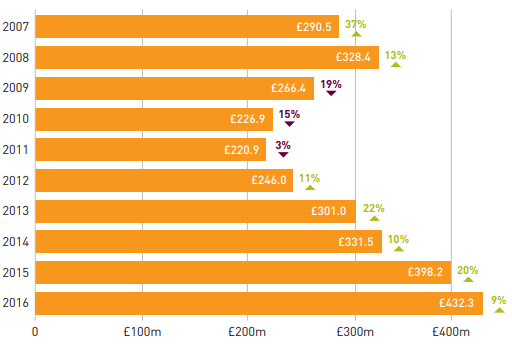}
    \caption{Increase in Remote purchase fraud losses on UK issued cards 
    \\Source: www.financialfraudaction.org.uk/fraudfacts17/assets/fraud\_the\_facts.pdf}
    \label{fig_CC_transactionFraud}
\end{figure}

Due to chip-based cards, counterfeit card-based fraud (using a duplicated copy of a stolen genuine card details) has declined (Figure~\ref{fig_CC_losses}). According to FFA, increasingly, credit card fraud appears to occur via online transactions and 2016 figures show a 20\% increase from the previous year UK numbers, as summarized in Figure~\ref{fig_CC_transactionFraud}.

It is evident, from Figure~\ref{fig_CC_transactionFraud}, that online card fraud is on the rise and its impact is increasing.  

Credit card frauds result in financial losses to all stakeholders, but ultimately hit the general public, who has to bear the higher bank charges and fees to buffer up the expenses and losses of the merchants and financial institutions. Reputation losses may affect the merchants and banks in the long run with the risk of customer switch over and instability. Mistrust and reluctance to use the payment systems can cause economic inefficiencies in terms of risk provisions, slowdown, and depression. Fraud incurs additional costs of card replacement, merchandise replacement, chargebacks and administration and staff overheads. 
According to US based LexisNexis Group (Lexisnexis.com,2018) the true loss on a merchant after remedial action was \$2.40 for each 1\$ fraud, in 2016, an 8\% increase over the year.

To avoid losses and damages from fraud two approaches, prevention and detection, can be used, often in unison. Fraud prevention is a more proactive method aimed at stopping frauds taking place by means of proper procedures, safe guards and controls. It minimises ability to conduct a fraud. Fraud detection is more reactive and comes in to play at each transaction that takes place in order to detect a fraudster in action. 

\subsection{Importance of credit card fraud detection}
According to \cite{zareapoor2015application}, fraud detection systems are essential to minimize losses to counter increasing frauds and fraud costs. While incidents of credit card fraud are limited to about 0.2\% of all card transactions, it may result in huge financial losses from large transaction values \cite{sohony2018ensemble}.
Since fraud cannot be prevented, given the human nature, early detection is used to minimize the damage. 

Fraud detection has been defined as the process that, given a set of credit card transactions, identifies if a new authorized transaction belongs to the class of fraud or normal transactions \cite{maes2002credit}. While such a process should be economical \cite{quah2008real} and practical in today’s institutional context and technology, it has also become mandatory for survival. While many research efforts are ongoing, progress is slow due to some unique challenges and the nature of the problem. 

\subsection{Challenges of Credit Card Fraud Detection}
Credit card fraud detection is extremely difficult \cite{duman2013solving} due to its volume, the adaptive and unique nature of each fraud and the need for real time or near real time assessments (requiring automated identification, classification, and annotation). Many researchers \cite{zareapoor2015application,sohony2018ensemble} state that constraints such as lack of real data sets due to sensitivity, confidentiality and privacy concerns and the massively imbalanced highly skewed nature of the data makes the problem challenging and difficult, if not impossible. 

\subsection{Machine learning}
Machine learning is a technique where a software applications \textit{learns} predict an outcome, without being explicitly programmed. This is accomplished by using historical data as input to predict new output values.
Machine learning techniques for fraud detection can be supervised, semi-supervised or unsupervised \cite{buczak2015survey}. In supervised learning, human annotated (labelled)  fraud and normal transactions are used the train a (software) model how to predict frauds. In the case of classification, the model determines if a given transaction is fraudulent; in the case of in case of regression it provide a risk rating. Aim is for investigators can prioritize on a smaller sub set of highly probable frauds. In unsupervised learning, the model attempts to cluster transactions in to fraud and normal groups based on similarities of features without the help of human pre-labelled data \cite{sathyapriya2017big}. Finally, in semi-supervised techniques, a partially labelled data set is used \cite{pise2008survey}. Generally most many machine learning models can be used in all modes, though with varying degree of accuracy and reliability, though supervised algorithms and models are most common \cite{duman2013novel}.

Some researchers \cite{zojaji2016survey} consider user behaviour a proxy for fraud detection, assuming that a frustrater's use of the card should deviate from normal, thus containing what is known as an anomaly, which stands out. This, also known as a one-class classification \cite{zheng2019one} has the benefit of only needing normal data for the learning process, which is more common and consists of nearly 99\% of the transactions \cite{duman2013solving}. However, while user behaviour analysis is able to detect novel frauds easily \cite{zojaji2016survey}, they have a  high rate of false alarms (FP). With current high transaction rates, this makes such models unattractive \cite{zojaji2016survey}. Further, sophisticated fraudsters deliberately emulate normal user patterns \cite{zheng2019one}; making void the different behaviour assumption on which these models are based upon. 

Unsupervised models require, while not requiring labelled data, needs excessive training and time to reach acceptable levels of performance \cite{zojaji2016survey}. Also, with large data volumes, such times and computational costs can reach impractical thresholds.  

After the model is developed and trained, its \textit{learning} can be stored, copied and reused on multiple and different systems \cite{buczak2015survey} for predictions.

\subsection{Context of the research}
Given a sufficiently detailed and large labelled collection of historical transactions, machine learning can be used to “learn” complex characteristics and flag frauds in new transactions, ideally in real-time. It will result in lesser number of frauds as cards could be blocked immediately on the very first instance preventing multiple frauds being carried out. It will also act as a deterrent and a reassurance to users. Card providers utilizing such technology would see return on investment (ROI) from increased customer volumes in the short term, while in the long term such mechanisms would be core competencies for survival.

Yet, research in this area is slow. In this work, we investigate the characteristics of fraud data and explore the challenges and complexities needed to be overcome for effective application of machine learning researchers to this problem. 

A significant challenge is posed by the sensitive nature of transaction data.
It is a typical and accepted practice to only provide researchers access to limited and (mostly) somewhat obfuscated and encoded data. Almost always, some processing is done at least to remove the personal identifiable elements. As such, PCA is a common format for release of such confidential data sets. However, no detailed study has been carried out to explore the effect of PCA obfuscation over a wide range of different machine learning classifiers. 

In this work we:
\begin{itemize}
    \item Investigate characteristics of typical fraud detection data sets
    \item Explore the effect of PCA obfuscation on confidential data and its impact on algorithmic performance of classifiers
\end{itemize}

\section{Characteristics of credit card data sets and challenges}
The process of creating a model for credit card fraud detection using labelled transaction data has several major constraints and challenges associated \cite{duman2013solving,zojaji2016survey}.

\begin{itemize}
\item \textbf{Massive imbalance in the data:}  as frauds are highly outnumbered by normal transactions, typically about 0.2\%. Importantly, as observed by \cite{dal2015calibrating}, it is the minority class (frauds) that is of interest and that the scarcity of these important frauds often hampers learning ability of the classifier often resulting in poor models.
\item \textbf{Unavailability of data sets:} as card companies cannot release these to the public domain or researchers due to confidentiality, privacy, secrecy and legal concerns. Though there is an abundance of machine learning algorithms and a keen interest, any research work in this direction will need real data to progress \cite{dal2015calibrating}. Further, \cite{duman2013solving} noted the use of synthetic data as an alternative and the absence or difficulty to access research and findings.
\item \textbf{Limited published work:} as researchers do not generally disclosed the features of the data or parameters used in classifiers \cite{sohony2018ensemble}. This could be due to funding institution restrictions. According to \cite{sahin2011detecting}, the development of new credit card fraud detection methods is severely hampered by the difficulty in exchanging ideas in fraud detection due to security and privacy concerns.
\item \textbf{Adaptive and innovative fraudulent behaviour:} Profiles of fraud and normal changes too often and keeps on evolving \cite{duman2013solving} requiring constant re-learning of the new patterns. There is also the possibility of past fraudulent transactions mistaken as valid and the algorithm learning these incorrect facts. New frauds can be crafted to mimic normal transactions and one miss can result in all similar cases being missed. According to \cite{sohony2018ensemble}, the problem can be too complex to differentiate even by human experts.
\item \textbf{Overlap and limited separability of the classes:} is a direct consequence of the above.  It has been observed \cite{sohony2018ensemble}, using DBSCAN, Gaussian mixed model, Binomial Logistic Regression, and k-means, that fraudulent transactions are not outliers (anomalies) and that they are not isolated or limited to a specific cluster but distributed uniformly.
\item \textbf{Lack of suitable evaluation metrics:} suited for massively unbalanced data, as traditional accuracy scores cannot be used. Unfortunately, \cite{zojaji2016survey} et al. (2016) notes the non-existence of a standard evaluation criterion or metrics for measuring and comparing the performance and quality of fraud detection systems. As consequence, it impossible to benchmark detection models and algorithms. Further, \cite{duman2013solving} points out the lack of benchmarking as a key issue. Researches are divided as to what constitutes the ideal metric for evaluation and what models yield better performance.
\item \textbf{Fraud detection cost:} must be economically acceptable and lower than the loss of fraud itself. With the large range of frauds and huge volumes, preventing a fraud of a few dollars does not make much sense (\cite{zojaji2016survey}, and such financial trade-offs should be included in the model and detection system design. 
\end{itemize}

An effective and viable machine learning solution needs to address all above factors effectively and efficiently in order to develop a suitable classifier.

\section{Choice of evaluation metrics}
In classification, one of the most important measurements is evaluation of the effectiveness and quality of the classifier, or the model’s ability to predict the correct classes of new information. Since, in credit card fraud data the frauds consist of a very minute fraction, traditional measures such as Accuracy cannot be relied upon \cite{barandela2003strategies}. 

As a simple illustration, given that there are 2 frauds for every 1000 transactions (the same as in the primary real-world data set), even a faulty model that always predict a transaction as normal would achieve an accuracy of (998 /1000 * 100) 99.8\%. While Accuracy score alone would make this an exceptional classifier, in the context of massively imbalanced data this evaluation metrics becomes useless. However, there is no agreed standardized evaluation metrics for typical massively unbalanced data sets.

\section{Incorporating human annotation errors with classification errors}
\begin{figure}
    \centering \includegraphics[width=1.00\columnwidth]{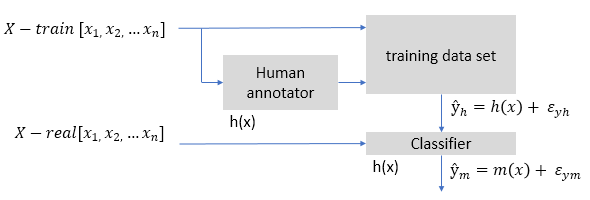}
    \caption{Formalizing the classification problem with real-world data}
    \label{fig_learningProblem}
\end{figure}

Let $X$ be the feature vector that contains a card transaction details. Further, $\textit{X-train}[x_1, x_2, \dots x_n]$ and $\textit{X-real}[x_1, x_2, \dots x_n]$ are respectively the train and test data sets. We use $h$ to denote human related operations and $m$ for machine related operations. Thus, $y, \hat{y}_h, \hat{y}_m$ are the real, human annotated and machine annotated labels and $\epsilon_{yh},\epsilon_{ym}$ are the human and machine error terms.

Thus, we have:
\begin{equation} 
\begin{aligned}
    \text{For the human annotation:} \quad 	& \hat{y}_h=h(x)+ \epsilon_yh \\
    \text{For the classification:} \quad    & \hat{y}_m=m(x)+ \epsilon_ym \\
\end{aligned}
\label{eq-1}
\end{equation} 

Also, since the classification model learns from X-train$[x_1, x_2, \dots x_n]$ and $\hat{y}_h$:
\begin{equation} 
    m = f(h, \epsilon_{yh})  
\label{eq-2}
\end{equation}

Therefore, for an arbitrary evaluation matrix of $S$:
\begin{equation} 
\begin{aligned}
    \text{Model error:} \quad 	& E_m = S( \hat{y}_h, \hat{y}_m) \\
    \text{Real error:} \quad   & E_r = S( y, \hat{y}_m) \\
\end{aligned}
\label{eq-3}
\end{equation} 

The best model can be obtained by minimization of the model error
\begin{equation} 
    E_m^* = \text{arg min}(E_m)  
\label{eq-3}
\end{equation}

However, the real-world requirement is minimization of the real error, which includes components of the human error and the model error:
\begin{equation} 
    E_r^* = \text{arg min}(E_r)  
\label{eq-4}
\end{equation}

Therefore, it is evident that while model accuracy is important, the human annotation accuracy plays a significant role in the final classifier accuracy which cannot be ignored.

Given that the human error falls beyond the scope of machine learning, it is still possible to apply some form of anomaly detection or clean up, purely to address the errors caused at the annotation stage and minimize $E_r$ with the expectation of making the classifier’s learning task easier and error rates low.

It can also be assumed that an error in the classification of any of the minority classes (false negatives, i.e.: misclassification of fraud as normal) can have a stronger impact on the final plateau of performance that a model is able to reach. It is also safe to assume that the level of imbalance has an effect on error rates as it increases the impact of a single error.

\section{Unbalanced classification}

According to \cite{duman2013solving} the fraud detection problem is difficult as there are relatively minute examples of fraud and massive imbalance.
Further, \cite{dal2015calibrating} states that such minute fraud fractions suffer from concept drift, where transactions might change their statistical properties over time.

Sampling is used to either lower the majority class (undersampling) or raise the minority (oversampling) to re-balance the data. Researchers \cite{dal2015calibrating} observed that balancing or leaving the training set unbalanced has a direct influence on the resultant model. While oversampling replicates the minority class giving the classifier enough opportunity to learn, it does not add any new information and may lead to overfitting the majority class especially in the case of noisy input \cite{sohony2018ensemble}. According to \cite{chawla2002smote}, oversampling increases training time and by creating copies of the minority (frauds) and is susceptible to overfitting.

Undersampling limits the majority class. While it leads to a significant loss of data, it creates a model on real observed data. However, undersampling modifies the priors and consequently biases the posterior probabilities of a classifier \cite{dal2015calibrating}. 

While oversampling replicates the minority class giving the classifier enough opportunity to learn, it does not add any new information and may lead to overfitting the majority class especially in the case of noisy input \cite{sohony2018ensemble}. It also increases training time \cite{chawla2002smote}. 

Notably, using oversampling or undersampling results in the training data set having a different distribution from the real data set. It results in a biased training set which is known in literature as Sample Selection Bias. 

\begin{table}[]
    \begin{tabular}{p{0.70\linewidth} | >{\centering\arraybackslash}p{0.30\textwidth} }
    \toprule
    Undersampling method                                                    & Reference             \\ \midrule
    Random majority under-sampling with   replacement                       &                       \\
    Instance Hardness Threshold                                             & Smith et. al. (2014) \cite{smith2014instance}  \\ \bottomrule
    \end{tabular}
    \caption{Undersampling methods in the in the imbalanced-learn python library}
    \label{TABLE_undersampleing}
\end{table}
\begin{table}[]
    \begin{tabular}{p{0.70\linewidth} | >{\centering\arraybackslash}p{0.30\textwidth} }
    \toprule
    Oversampling method                                                     & Reference             \\ \midrule
    Random minority over-sampling with   replacement                        &                       \\
    SMOTE - Synthetic Minority Over-sampling   Technique                    & Chawla et. al. (2002) \cite{chawla2002smote} \\
    ADASYN - Adaptive synthetic sampling approach for imbalanced learning & He et al. (2008) \cite{he2008adasyn}    \\ \bottomrule
    \end{tabular}
    \caption{Oversampling methods in the in the imbalanced-learn python library}
    \label{TABLE_oversampleing}
\end{table}

Table~\ref{TABLE_undersampleing} shows some of the undersampling methods implemented in the imbalanced-learn python library \footnote{Contrib.scikit-learn.org, 2018} which is used later in this work for experiments and classifier selection. The imbalanced-learn python library also provides oversampling methods, which are given in Table~\ref{TABLE_oversampleing}. 

\section{Access to training data sets}
Researchers need to make a trade-off between the real world and practical data, due to constraints such as privacy and confidentiality, especially in domains as credit card transactions.
One of the biggest challenges and a major hurdle in research related to credit card fraud detection is the lack of real-world data sets \cite{duman2013solving}.  Notably, \cite{zojaji2016survey} states that "the lack of publicly available database has been a limiting factor for the publications on financial fraud detection” and notes the unavailability of a universally accepted field structure for credit card transactions. A sample of the observations by \cite{duman2013solving}, in Table~\ref{TABLE_CC_data sets} summarizes the wide spectrum and diversity of the fraud distributions. Note that these are data sets released by respective sources, \textbf{after} undergoing possible vetting. Further, they can be from differing contexts.

\begin{table}[]
    \setlength{\tabcolsep}{5pt}
    \begin{tabular}{
      >{\centering}p{0.3\textwidth}
      >{\centering}p{0.4\textwidth}
      >{\centering\arraybackslash}p{0.3\textwidth}
    }
    \toprule
    Source   & Description & Researcher \\ \midrule
    Large Brazilian bank   (Real Data set)   & 41647 transactions,   3.14\% frauds     & Manoel Fernando, Alonso   Gadi et al. (2008)  \\ \midrule
    Financial institute in   Ireland (WebBiz)(Real Data set) & 4 million transactions, 0.13\% frauds  & Anthony Brabazon et al.   (2010)    \\ \midrule
    Hong kong bank (Real   Data set)         & 50 million transactions    & C. Paasch (2007), Siddhartha Bhattacharyya et al.   (2010) \\ \midrule
    Chase Bank and First   Union Bank Real Data set)       & Chase Bank: 500,000   transactions, 20\% frauds. First Union: 500,000   transactions, 15\% frauds & Philip K. Chan (1999)  \\ \midrule
    Major US bank (Real   Data set)         & 6000 transactions, 16\%   frauds, 64 features   & G. Kou, et al. (2005)  \\ \midrule
    Large Australian bank   (Real Data set)  & 640361 transactions    & Nicholas Wong et al.   (2012)  \\ \midrule
    Vesta Corporation (Real   Data set)      & 206,541 transactions,   1.2\% fraud    & John Zhong Lei (2012)    \\ \midrule
    Mellon Bank (Real Data set)              & 1million transactions   & SushmitoGhosh(1994)    \\ \midrule
    Synthetically generated   data           & 320million   transactions, 42 features        & M. Hamdiozcelik et al.   (2010)     \\ \midrule
    Synthetically generated   data           & 1000000 transactions, 20 features    & K.RamaKalyani et al.   (2012)    \\ \midrule
    Synthetically generated   data           & 10000 transactions   & Tao guo et al. (2008)   \\ \midrule
    Synthetically generated   data           &       & MubeenaSyeda et al.   (2002)    \\ \bottomrule                   
    \end{tabular}
    \caption{Credit Card fraud Data sets used by previous Researchers, originally from \cite{zojaji2016survey}, modified by author}
    \label{TABLE_CC_data sets}
\end{table}

Credit card transaction data sets used by researchers reveals that:
\begin{itemize}
    \item \textit{Fraud percentage in every data set is different ( and some seem to be artificially adjusted; ex: 80-20 for public release)} 
\end{itemize}
While this can be explained by the different fraud prevention mechanisms and deterrents in place \cite{zojaji2016survey}, or the preference to hide the true fraud rates, it indicates that the credit card fraud characteristics are unique across countries and institutions. The only similarity remains to be the massively unbalanced nature. 
Therefor it seems a model selection strategy based on the data set, rather than a generally effective model, is more suitable. 

\begin{itemize}
    \item \textit{Number of features used is different and it is safe to assume that they would have different assumptions and definitions.}
\end{itemize}
This implies that a solution for credit card detection should not place too much emphasis on data exploration and engineering. Given that most information is confidential it may not even be possible to know the definitions of certain fields or its content in \textit{raw} form.

\section{Case study : the Pozzolo data set}
Credit card information and data sets are confidential in nature and while companies are keen to research and break new frontiers in machine learning and fraud detection, they are also reluctant to hand over sensitive customer information to the academia. In addition to the ethical and public relations issues the is also legal constraints in most parts of the world. This has created a huge problem in the research efforts. 
Due to the lack of real-world data, researchers tend to use a few such publicly available data sets, often released with personal information obfuscated and/or de-identified for legal, ethical, and business reasons. 

Principal Component Analysis (PCA) is commonly used to encode and obfuscate content.  The principal components of a collection of data points are a sequence of $p$ unit vectors, where the $i$-th vector is the direction of a line that best fits the data while being orthogonal to the first $i-1$ vectors. Given the principal components, it is not possible to recover the original data. 

PCA has the following characteristics:
\begin{itemize}
    \item Reduces the dimensionality of the data set, lowering the CPU, memory and time
    \item Ensures that personal and confidential information remains safe
    \item Removes the dependants on fields and variables.
\end{itemize}

The PCA encoded real-world credit card fraud data set \cite{dal2015calibrating} is such an example, where all features are PCA encoded prior to release and the actual features are unknown. 

The real-world credit card transaction data set, originally published by \cite{dal2015calibrating}, consists of 284,807 PCA encoded labelled transactions of 2 days in September 2013 by European cardholders. Of these, 492 or 0.172\% are frauds, indicating massively unbalanced data. It is PCA encoded to ensure confidentiality and consists of 28 principal components. It is released to public domain by the ULB Machine Learning Group (MLG), a research unit of the Computer Science Department of the Faculty of Sciences, Université libre de Bruxelles.  Table~\ref{TABLE_PCA_primary_dataset} summarizes its features. 

\begin{table}[]
    \begin{tabular}{p{0.20\linewidth} p{0.20\linewidth} p{0.40\linewidth} p{0.20\linewidth}}
    \toprule
    Normal  & Frauds & Features            & Instances \\ \midrule
    284,315 & 492    & 30 (28 PCA + 2 Raw) & 284, 807  \\ \bottomrule
    \end{tabular}
    \caption{Primary PCA encoded (Pozzolo) Data set }
    \label{TABLE_PCA_primary_dataset}
\end{table}

It should also be noted that PCA encoding does not necessitate (or allow) any feature engineering.

\section{Algorithmic efficiencies of PCA obfuscation on data}
In this section, to validate that assumption that PCA does not obscure or lose significant information and still allows models to achieve acceptable accuracy, we conduct a set of experiments. We use a secondary PCA un-encoded data set and compare classification accuracy of a wide range of popular machine learning classifiers with it and its PCA-encoded version. By comparison of the classification performance, we determine the effect of the PCA encoding on the algorithmic efficiency of classification in general.

Specifically, the aim is to determine if the primary data set still yields appropriate results despite the PCA obfuscation and importantly, to determine the effectiveness of PCA as a means of disseminating confidential data sets for machine learning research.

\subsection{The secondary data set}
The secondary data set, summarized in Table~\ref{TABLE_secondary_data set}, originally used by \cite{yeh2009comparisons} for binary classifier evaluation, is available from the University of California Irvine (UCI) Machine learning repository \footnote{Archive.ics.uci.edu, 2018}, and consists of raw (non PCA encoded) records related to client information, payment history, status and credit limits of credit card customers in Taiwan from April 2005 to September 2005 \cite{lichman2013uci}. 
\begin{table}[]
    \begin{tabular}{p{0.20\linewidth} p{0.20\linewidth} p{0.40\linewidth} p{0.20\linewidth}}
    \toprule
    Normal  & Frauds & Features  & Instances \\ \midrule
    23,364  & 6634   & 23 Raw   & 30,000   \\ \bottomrule
    \end{tabular}
    \caption{Secondary Data set }
    \label{TABLE_secondary_data set}
\end{table}

\begin{figure}
  \centering
  \subfloat[2\% frauds]{\includegraphics[width=0.48\textwidth]{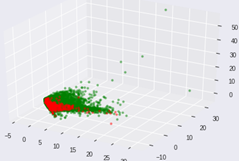}\label{fig_frauds}}
  \hfill
  \subfloat[22\% frauds]{\includegraphics[width=0.48\textwidth]{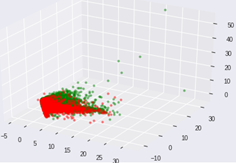}\label{fig:f2}}
  \caption{3D plot of the PCA encoded secondary data set with frauds}
  \label{fig_2_22}
\end{figure}

The secondary data has a high (22\%) fraud percentage and is less imbalanced than the primary Pozzolo data set. Thus, we adjust it to obtain a fraud rate of 2\%, by removing surplus frauds. Lowering the fraud rate beyond 2\% resulted in too few samples being present for the classifiers to work with which we avoid. The 3D PCA plots for full and reduced data sets are shown in Figure~\ref{fig_2_22}. There does not appear to be any noticeable difference in the 2 plots, other than the lesser number of frauds present in the leftmost 2\% model.

\subsection{Methodology}
To assess the effect of PCA encoding on classifier performance, we test the secondary data set with and without PCA encoding on classification. We test both the original 22\% and the 2\% modified variant, and report them separately, as the real-world data sets are often bellow 1\% fraud percentages and also given that there is a wide range of variation in fraud percentages of typical data sets. 
We crate a 23-principle component PCA encoded version of the secondary data set using the scikit library, specifically full SVD with a standard LAPACK solver.

A StratifiedShuffleSplit (i.e.: class proportion in the data sets is kept constant for all folds) with a 20:80 test: train split was used to train 15 machine learning classifiers from the scikik-learn library, specifically DummyClassifier, LogisticRegression, RandomForestClassifier, GaussianNB, SVC(Linear), MLPClassifier, RidgeClassifier, DecisionTreeClassifier, SGDClassifier, PassiveAggressiveClassifier, Perceptron, KneighborsClassifier and ensembles AdaBoostClassifier (Real), AdaBoostClassifier (Discrete), GradientBoostingClassifier along with QuadraticDiscriminantAnalysis. We use  F1 score and g-mean as effective performance measures for unbalanced data sets following \cite{dal2015calibrating}. 

Python is used as the main programming language with Scikit-learn and imbalanced-learn libraries. All code and models were run on Google Colaboratory with data stored on Google Drive.

\subsection{Experimental results and analysis}
\subsubsection{Effect of PCA on classifier's algorithmic efficiencies} 
\begin{figure}
    \centering \includegraphics[width=1.00\columnwidth]{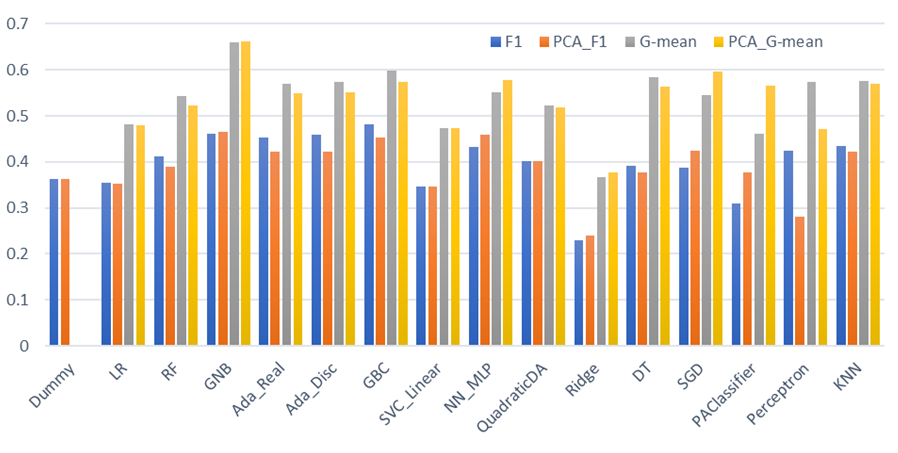}
    \caption{Raw vs PCA encoded – secondary data set - 22\% frauds}
    \label{fig_results22}
\end{figure}

\begin{figure}
    \centering \includegraphics[width=1.00\columnwidth]{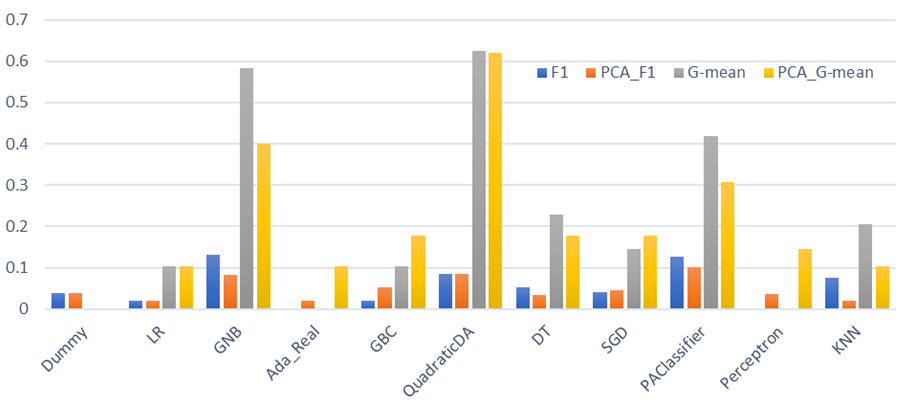}
    \caption{Raw vs PCA encoded – secondary data set - 2\% frauds}
    \label{fig_results_2}
\end{figure}

Figure~\ref{fig_results22} shows the effect of raw vs PCA encoded original (22\% fraud) secondary data set. It can be seen that in case of LR, GNB, SVC and QuadraticDA the F1 scores are identical. The g-mean is also quite similar. In Ridge, SGD and PAClassifier the combined information content in PCA has resulted in a high F1 and a higher g-mean score, indicating a quite good information survival during the PCA process. The cases of RF, Ada and KNN where F1 and g-mean are lower are still within acceptable tolerances. This indicates that for semi balanced data sets with a minority of fraud, PCA seems to generally preserve the informational content across binary classifiers for F1 and g-mean evaluation metrics.

Figure~\ref{fig_results_2} shows the adjusted (2\%, fraud rate) secondary data set results. With the lower frauds and a smaller data set, some of the models were not able to calculate well defined F1 values and were removed from the graph.  The F1 scores are similar for LR, SGD and QuadraticDA. GNB, DT and KNN shows degraded F1 while GBC shows improved scores. 

We observe that PCA does not preclude the learning and information content in massively unbalanced data sets. However, learning is more challenged with increasing imbalance. Also, F1 and g-mean act as independent scores as there is little correlation between these.

\subsubsection{Effect of dimensions on classifier's algorithmic efficiencies} 
\begin{figure}
    \centering \includegraphics[width=1.00\columnwidth]{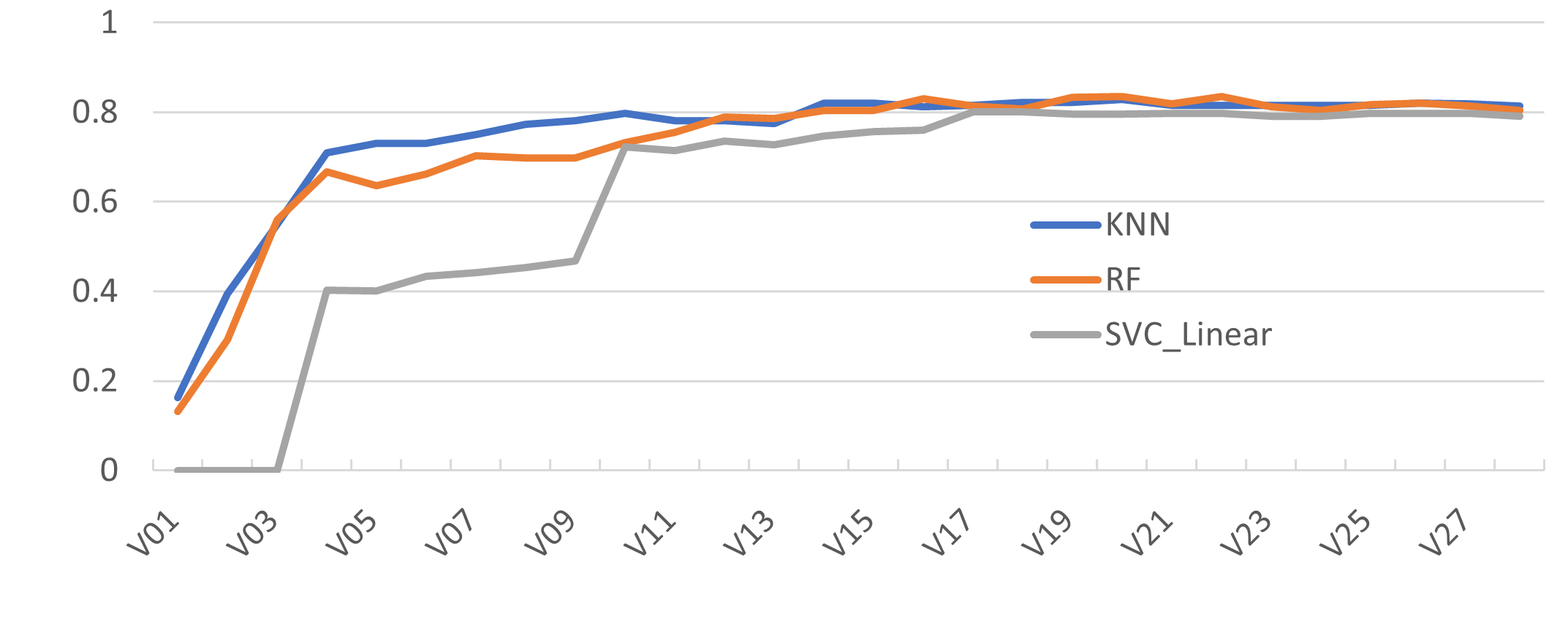}
    \caption{Effect of Dimensions on Classifier performance}
    \label{fig_results_DIM}
\end{figure}

Figure~\ref{fig_results_DIM} shows the change of F1 score based on the dimensionality for the top 3 models. They peek around the first 15 principle components. Thereafter, minor degradation on the F1 score is visible, possibly due to overfitting in the presence of a strong negative class majority. 

Therefore, while PCA does not significantly degrade performance, care should be taken to use the appropriate principle component size (dimensions) to avoid overfitting.

\subsubsection{In summary,} these experiments demonstrate that PCA does not preclude learning in massively unbalanced data sets. 
PCA obfuscation preserves adequate information content and does not have significant impact on the classifier performance. All classifiers were able to report near-identical scores both at moderate and massive imbalance scenarios. However there is significant reduction in learning with increasing imbalance. Specifically, learning is more challenging mainly due to the unbalanced nature and not PCA obfuscation.

\section{Limitations}
Due to unavailability of a data set, this research was unable to look in to the effect of a less low than 1\% imbalance. The secondary date set used was only able to be trimmed to 2\%. There is scope for a more in-depth analysis of the effectiveness of PCA and also LDA as an alternative for obfuscation.

Further evaluation of transaction-based classification cost models (as done by few researchers) along with a larger spread of classifiers. It would necessitate base classifier algorithms to be modified to incorporate custom cost models and would be worthwhile investigating as many researchers have already suggested novel classification cost models including class based, transaction based and even item based (e.g.: the remaining balance value on the credit card).

\section{Conclusion}
In this work we provide descriptive and experimental insights in to the properties of credit card fraud data, focusing on its nature, challenges and implications of obfuscation as a means to ensure confidentiality). We investigate the algorithmic performance of 15 machine learning classifiers against PCA encoded sensitive transaction data and show that, while PCA does not significantly degrade performance, care should be taken to use the appropriate principle component size (dimensions) to avoid overfitting.
Furthermore, we show how human annotation errors compound with machine classification errors.

In addition to a severe lack of suitable machine learning data sets for fraud detection research, of those available, each credit card data set gives conflicting characteristics (feature definitions, depth and breadth) as well as fraud ratios which could be the cause of the inability to obtain an ideal general classification algorithm. Further, there is no standardized corpus that defines what should be typical credit card data, despite the universal and ubiquitous nature of global card payments. The only similarity remains to be the massively unbalanced nature. Therefore, it may not be possible to find a one-hat solution for credit card fraud detection and a data-driven approach is necessary.

\section{Acknowledgments}
Dedicated to Sugandi.

\bibliographystyle{splncs04}
\bibliography{main}

\begin{thebibliography}{10}
\providecommand{\url}[1]{\texttt{#1}}
\providecommand{\urlprefix}{URL }
\providecommand{\doi}[1]{https://doi.org/#1}

\bibitem{chawla2002smote}
Smote: synthetic minority over-sampling technique. Journal of artificial
  intelligence research  \textbf{16},  321--357 (2002)

\bibitem{barandela2003strategies}
Barandela, R., Sanchez, J., Garcia, V., Rangel, E.: Strategies for learning in
  class imbalance problems. Pattern Recognition  \textbf{3}(36),  849--851
  (2003)

\bibitem{buczak2015survey}
Buczak, A.L., Guven, E.: A survey of data mining and machine learning methods
  for cyber security intrusion detection. IEEE Communications surveys \&
  tutorials  \textbf{18}(2),  1153--1176 (2015)

\bibitem{dal2015calibrating}
Dal~Pozzolo, A., Caelen, O., Johnson, R.A., Bontempi, G.: Calibrating
  probability with undersampling for unbalanced classification. In: 2015 IEEE
  symposium series on computational intelligence. pp. 159--166. IEEE (2015)

\bibitem{duman2013novel}
Duman, E., Buyukkaya, A., Elikucuk, I.: A novel and successful credit card
  fraud detection system implemented in a turkish bank. In: 2013 IEEE 13th
  International Conference on Data Mining Workshops. pp. 162--171. IEEE (2013)

\bibitem{duman2013solving}
Duman, E., Elikucuk, I.: Solving credit card fraud detection problem by the new
  metaheuristics migrating birds optimization. In: International
  Work-Conference on Artificial Neural Networks. pp. 62--71. Springer (2013)

\bibitem{he2008adasyn}
He, H., Bai, Y., Garcia, E.A., Li, S.: Adasyn: Adaptive synthetic sampling
  approach for imbalanced learning. In: 2008 IEEE international joint
  conference on neural networks (IEEE world congress on computational
  intelligence). pp. 1322--1328. IEEE (2008)

\bibitem{lichman2013uci}
Lichman, M., et~al.: Uci machine learning repository (2013)

\bibitem{maes2002credit}
Maes, S., Tuyls, K., Vanschoenwinkel, B., Manderick, B.: Credit card fraud
  detection using bayesian and neural networks. In: Proceedings of the 1st
  international naiso congress on neuro fuzzy technologies. vol.~261, p.~270
  (2002)

\bibitem{pise2008survey}
Pise, N.N., Kulkarni, P.: A survey of semi-supervised learning methods. In:
  2008 International conference on computational intelligence and security.
  vol.~2, pp. 30--34. IEEE (2008)

\bibitem{quah2008real}
Quah, J.T., Sriganesh, M.: Real-time credit card fraud detection using
  computational intelligence. Expert systems with applications  \textbf{35}(4),
   1721--1732 (2008)

\bibitem{sahin2011detecting}
Sahin, Y., Duman, E.: Detecting credit card fraud by ann and logistic
  regression. In: 2011 international symposium on innovations in intelligent
  systems and applications. pp. 315--319. IEEE (2011)

\bibitem{sathyapriya2017big}
Sathyapriya, M., Thiagarasu, V.: Big data analytics techniques for credit card
  fraud detection: A review. International Journal of Science and Research
  \textbf{6}(5),  206--211 (2017)

\bibitem{smith2014instance}
Smith, M.R., Martinez, T., Giraud-Carrier, C.: An instance level analysis of
  data complexity. Machine learning  \textbf{95}(2),  225--256 (2014)

\bibitem{sohony2018ensemble}
Sohony, I., Pratap, R., Nambiar, U.: Ensemble learning for credit card fraud
  detection. In: Proceedings of the ACM India Joint International Conference on
  Data Science and Management of Data. pp. 289--294 (2018)

\bibitem{yeh2009comparisons}
Yeh, I.C., Lien, C.h.: The comparisons of data mining techniques for the
  predictive accuracy of probability of default of credit card clients. Expert
  systems with applications  \textbf{36}(2),  2473--2480 (2009)

\bibitem{zareapoor2015application}
Zareapoor, M., Shamsolmoali, P., et~al.: Application of credit card fraud
  detection: Based on bagging ensemble classifier. Procedia computer science
  \textbf{48}(2015),  679--685 (2015)

\bibitem{zheng2019one}
Zheng, P., Yuan, S., Wu, X., Li, J., Lu, A.: One-class adversarial nets for
  fraud detection. In: Proceedings of the AAAI Conference on Artificial
  Intelligence. vol.~33, pp. 1286--1293 (2019)

\bibitem{zojaji2016survey}
Zojaji, Z., Atani, R.E., Monadjemi, A.H., et~al.: A survey of credit card fraud
  detection techniques: data and technique oriented perspective. arXiv preprint
  arXiv:1611.06439  (2016)

\end{thebibliography}
\end{document}